\documentclass{article}
\usepackage[margin=1.5in]{geometry}
\usepackage[hyphens]{url}
\usepackage{graphicx}
\usepackage[round]{natbib}
\usepackage[colorlinks=true, allcolors=blue]{hyperref}
\usepackage{amsmath}
\usepackage{amssymb}
\usepackage{booktabs}
\usepackage{url}
\usepackage[T1]{fontenc}
\usepackage{multirow}

\newcommand{\dataset}[1]{{\textbf{\texttt{#1}}}}
\newcommand{\domain}[1]{\textit{#1}}
\newcommand{\name}[1]{DomainDemo}

\title{\name{}: a dataset of domain-sharing activities among different demographic groups on Twitter}

\author{
    Kai-Cheng Yang,\textsuperscript{\rm 1}
    Pranav Goel,\textsuperscript{\rm 1}
    Alexi Quintana-Math\'{e},\textsuperscript{\rm 1}
    Luke Horgan,\textsuperscript{\rm 1}\\
    Stefan D. McCabe,\textsuperscript{\rm 1}
    Nir Grinberg,\textsuperscript{\rm 2}
    Kenneth Joseph,\textsuperscript{\rm 3}
    and David Lazer\textsuperscript{\rm 1}\\
    \\
    \textsuperscript{\rm 1} Network Science Institute, Northeastern University, Boston, MA, USA \\
    \textsuperscript{\rm 2} Department of Software and Information Systems Engineering,\\ Ben-Gurion University of the Negev, Be'er-Sheva, Israel \\
    \textsuperscript{\rm 3} Computer Science and Engineering Department,\\University at Buffalo, Buffalo, NY, USA
}

\begin{document}
\maketitle

\begin{abstract}
Social media play a pivotal role in disseminating web content, particularly during elections, yet our understanding of the association between demographic factors and information sharing online remains limited.
Here, we introduce a unique dataset, DomainDemo, linking domains shared on Twitter (X) with the demographic characteristics of associated users, including age, gender, race, political affiliation, and geolocation, from 2011 to 2022.
This new resource was derived from a panel of over 1.5 million Twitter users matched against their U.S. voter registration records, facilitating a better understanding of a decade of information flows on one of the most prominent social media platforms and trends in political and public discourse among registered U.S. voters from different sociodemographic groups.
By aggregating user demographic information onto the domains, we derive five metrics that provide critical insights into over 129,000 websites.
In particular, the localness and partisan audience metrics quantify the domains' geographical reach and ideological orientation, respectively.
These metrics show substantial agreement with existing classifications, suggesting the effectiveness and reliability of DomainDemo's approach.
\end{abstract}

\section*{Background \& Summary}

Social media play a significant role in the distribution of information~\citep{kietzmann2011social,osatuyi2013information,gaal2015exploring}, serving as essential sources of information for millions of users, especially in critical contexts such as elections~\citep{himelboim2013playing,holt2013age} and public health crises~\citep{gui2017understanding,kudchadkar2020using,mitchell2020americans,mitchell2021americans}.
The content shared on social media originates not only from the users themselves but also from a wide array of sources.
In particular, posting and re-sharing links to external websites (URLs) are key mechanisms for disseminating web content on social media~\citep{kwak2010twitter,ju2014will,pew2023social,rosenstiel2015twitter}.
For some websites, this is a crucial way of getting traffic and for users to access their content~\citep{cordrey2012tanya,phillips2012sociability,newman2016overview,myllylahti2018attention}.
For instance, \citet{chen2022we} demonstrate that social media are responsible for a substantial portion of referred visits to \domain{thegatewaypundit.com}, a popular far-right news website, by analyzing its web traffic data.

The significance of social media in information distribution has attracted considerable research attention in recent years.
However, our understanding of the role demographic factors play in this process remains limited~\citep{golder2014digital,shugars2021pandemics}, even though their importance in online discourse has been discussed in prior work~\citep{allen2013online,ting2017predicting,falenska2024self,barbera2015understanding}.
Typically, the data provided by platforms to researchers lack user-level demographic information~\citep{veale2017fairer,holstein2019improving,andrus2021we}.
When researchers rely on user donations or surveys to collect demographic data, the sample size is often insufficient to provide meaningful aggregate insights about content-sharing patterns, particularly at the domain level.
These challenges in data collection have created a significant gap in our ability to comprehensively analyze the interplay between demographic factors and online political discourse.

To bridge this gap, we introduce a novel dataset, \name{}, which quantifies domain-sharing activities across diverse demographic groups on Twitter.
Although Twitter was rebranded as X in 2023, we will refer to the platform as Twitter in this article since our dataset predates the rebranding.
Our data encompasses demographic details such as age, gender, race, political affiliation, and geolocation (U.S. state).
These domain-sharing events are derived from a comprehensive dataset of over 1.5 million U.S.-based Twitter users matched with their voter registration records.
Spanning 132 months (11 years) from May 2011 to April 2022, the dataset is organized in monthly intervals, enabling the analysis of temporal trends.
Our released datasets allow researchers to investigate the association between demographic characteristics and the sharing patterns of diverse information sources, ranging from mainstream news websites to potential sources of misinformation.

We release two versions of the domain-sharing statistics dataset: \name{}-multivariate and \name{}-univariate.
\name{}-multivariate splits the statistics into buckets defined by age, gender, race, political affiliation, state, and time altogether.
In each bucket, we provide the number of shares, the number of unique users sharing the domain, and the Gini index~\citep{gini1936measure} of the sharing count among users.
\name{}-univariate includes five universes: state, race, gender, age, and political affiliation.
Within each universe, we provide the number of shares, the number of unique users sharing the domains, and the Gini index of the sharing count among users for each category (e.g., age groups in the age universe).
In addition to the domain-level statistics, we also provide the distribution of the population in different demographic buckets for both \name{}-multivariate and \name{}-univariate.
These population-level distributions can serve as baselines.

\name{}-multivariate is similar in format to the Facebook URL dataset shared through Social Science One~\citep{king2020new,fburl}, which includes the number of views, shares, and reactions to various URLs across different demographic groups.
However, there are some key differences.
\name{}-multivariate focuses solely on a sample of Twitter users who are registered voters in the U.S., while the Facebook URL dataset includes data from all eligible users on the platform.
Additionally, \name{}-multivariate only includes the number of shares and unique users at the domain level and does not incorporate any noise.
However, our dataset provides more detailed demographic information compared to the Facebook URL dataset.
Specifically, while the Facebook URL dataset only includes country-level geographic data, our dataset contains state-level geolocation information and includes race information.

In addition to the count statistics, we introduce five derived metrics that quantify how the demographics (age, gender, race, political affiliation, and geolocation) of users sharing a particular domain differ from that of the baseline.
These metrics have specific interpretations that are useful for various research questions.
For instance, the geolocation of the users allows us to measure the domains' localness, i.e., the extent to which the sharing of a particular domain is geographically localized.
Together with the user-sharing behavior data, the localness metric enables researchers to quantify the changing landscape of the local news industry.
As a fundamental component of the U.S. democratic process, local news is uniquely positioned to report on local affairs and elections~\citep{martin2019local,moskowitz2021local}, but faces a declining trend over the years~\citep{hayes2018decline,abernathy2018expanding}.
Similarly, the user party affiliation in \name{} allows us to measure the audience partisanship of different domains.
This metric can serve as a proxy for the political leaning of the domains, crucial for understanding online political discourse~\citep{bakshy2015exposure,robertson2023users}.
Our derived metrics demonstrate strong alignment with established measures of localness and political leaning while significantly expanding coverage to over 129,000 domains---over ten times the number of domains in existing datasets.
The metrics also uncover subtle variations in sharing patterns that previous binary or one-dimensional categorization schemes could not capture.

Due to the difficulty of obtaining data from social media, especially Twitter, in the post-API era~\citep{murtfeldt2024riptwitterapieulogy}, replicating our efforts is challenging.
Even if access to Twitter data were to become available in the future, the platform itself has undergone significant changes.
These factors make our dataset a unique and valuable contribution to the research community, as it provides a comprehensive view of domain-sharing behaviors across an 11-year period.

\section*{Methods}

In this section, we describe how we create our dataset and provide case studies to interpret the data.

\subsection*{Twitter Panel}

Our dataset is based on a panel of over 1.5 million registered U.S. voters on Twitter, created by our team in previous work.
A pilot version of the panel was first used by \citet{grinberg2019fake}, then the panel was expanded considerably by ~\citet{shugars2021pandemics} and validated by \citet{hughes2021using}.
To create the panel, we start with the Twitter Decahose, a 10\% random sample of all tweets, and identify 290 million accounts that post content between January 2014 and March 2017.
We extract the names of the users, either from the Twitter handles or display names, and their location from the account profiles.
This information is then matched against voter data provided by TargetSmart in October 2017, covering all 50 U.S. states and the District of Columbia.
We compare the full name of each person in the voter file with the names of the Twitter accounts.
If the full name has fewer than 10 exact matches, we then examine the location of the Twitter accounts.
A Twitter account and voter record pair is accepted only if that is the only person in the specified city or state-level geographic area in both datasets.
This reliance on full names and disclosed locations helps to eliminate many fake, automated (bot), and organizational accounts.

The data collection and matching of the Twitter panel were approved by the Northeastern University Institutional Review Board (protocol number: 17-12-13).
Following the best practices outlined by \citet{hemphill2022comparative}, we employ data aggregation, anonymization, and access control measures to protect user privacy and minimize the risk of re-identification in our Twitter panel.

Matching to voter file records provides access to the geolocation, year of birth, gender, race/ethnicity, and partisanship of the users in the panel.
We use state-level geolocation data from the voter files as our geographic unit of analysis.
While we have access to more detailed location information, such as county-level data, releasing this information would risk re-identifying the users due to the low population density of many U.S. counties.
State-level granularity offers a good compromise between the usefulness of the data and the privacy of the users.
Using the year of birth, we determine the age of users at the time of sharing events and categorize them into the following age groups: ``<18,'' ``18-29,'' ``30-49,'' ``50-64,'' and ``65+.''
The category for users younger than 18 years old is included because some states allow 17-year-olds to pre-register to vote, and some users might be younger than 18 at the time of sharing events.
Gender is a binary measure provided by TargetSmart, which does not capture gender identities beyond the binary framework~\citep{shugarsgender}.
Race/ethnicity information is inferred by TargetSmart for most states and is categorized as ``African-American,'' ``Asian,'' ``Hispanic,'' and ``Caucasian.''
Other race categories with limited representation in the dataset are aggregated into a single ``Other'' category to minimize re-identification risks.

TargetSmart provides two measures of partisanship: party registration and inferred partisanship.
Party registration information in voter files is self-reported and aligns well with survey self-reporting~\citep{Ansolabehere_Hersh_2017}.
However, this information is unavailable for 20 states (AL, AR, GA, HI, IL, IN, MI, MN, MO, MS, MT, ND, OH, SC, TN, TX, VA, VT, WA, and WI) in the TargetSmart data, which account for 42.7\% of the Twitter users in our panel.
When categorizing party registration information, we treat values for users in the 20 aforementioned states as missing.
For the other 30 states and the District of Columbia, users registered as ``Democrat'' and ``Republican'' are coded accordingly.
Due to variations in the classification of independent registered voters by state, we group individuals listed as ``Independent,'' ``No party,'' or ``Unaffiliated'' into a single ``Independent'' category.
Members of minor parties, such as the Green Party and the Libertarian Party, are categorized as ``Other.''

Based on party registration and other indicators, TargetSmart infers the probability of all individuals in all 50 states and the District of Columbia voting Democrat.
We categorize individuals as Republican (0-0.35), Independent (0.35-0.65), and Democrat (0.65-1) using TargetSmart's recommended thresholds to generate the inferred partisanship.
For our data release and analysis, we use inferred partisanship as the primary measure since it covers all users (referred to as ``party'' hereafter).
Additionally, we provide party registration information as a secondary measure (referred to as ``party registration'' or ``partyreg'' hereafter), as it conveys a slightly different signal and offers useful insights for certain analyses.

Missing values in all dimensions are coded as ``Unknown.''

\subsection*{Domain-sharing Statistics}

We collect posts from users in the panel spanning from May 2011 to April 2022.
We extract the links shared by these users, expand the shortened links when possible, and identify the corresponding domains (e.g., \domain{nytimes.com} for The New York Times).
This process allows us to determine which user shares what domains and when.
Sharing events, as defined in our study, include posting links in original tweets and retweeting or quoting tweets containing links.
To reduce noise and the risks of re-identification, we include only domains that are shared by at least 50 unique users throughout the entire period.

We integrate the demographic information of users with their domain-sharing records to construct a comprehensive table.
This domain-sharing event table includes the following columns: user\_id, domain, age, gender, race, party, party registration, state, and year-month.
Each row corresponds to a single sharing event, with users who share the same domains multiple times contributing multiple rows.
Due to the presence of user identifiers, we cannot release this detailed table.
Instead, we provide aggregate statistics derived from this table including \name{}-multivariate and \name{}-univariate.

\name{}-multivariate describes the domain sharing behavior of users across different demographic dimensions simultaneously.
It includes several variants designed to facilitate different types of research analyses.
The most granular variant is the monthly distribution data at the domain level, which is produced by grouping the domain-sharing event table by domain, age, gender, race, party (party registration is excluded here), state, and year-month.
In each bucket, we calculate the following statistics: the number of shares, the number of unique users who share the domain, and the Gini index of the sharing count across users.
Formally, the Gini index $G$ for a domain is calculated as:
\begin{align}
    G &= \frac{1}{2 N^2 \bar{x}} \sum_{i=1}^N \sum_{j=1}^N |x_i - x_j| \text{,}
\end{align}
where $N$ is the number of users who share the domain, $x_\alpha$ is the number of shares by the $\alpha$-th user, and $\bar{x} = \sum_{i=1}^N x_i / N$ is the average number of shares per user.
$G$ ranges from 0 to 1, with 0 indicating equal sharing and values close to 1 indicating that a few users share the domain disproportionately.
Note that we set the Gini index to 1 for domains shared by only one user in the bucket.
We include the Gini index to help researchers understand the inequality of sharing events across users without releasing detailed information about these users due to privacy concerns.

In addition to the demographic distribution of users sharing each domain, it is useful to understand the distribution of the whole population in many cases.
Therefore, we also include a ``baseline'' variant that groups the domain-sharing event table by age, gender, race, party, state, and year-month.
In each bucket, we calculate the same statistics as the monthly distribution data.
We also provide the average number of unique domains shared by users in each demographic bucket and the corresponding standard deviation.
On top of the monthly data, we further provide the distribution and baseline data covering the whole time period.
In total, \name{}-multivariate includes four variants.

\name{}-univariate is generated by aggregating the sharing events across all demographic dimensions except for the one of interest.
For example, the state univariate data (referred to as the ``state universe'') is produced by aggregating the sharing events across all age, gender, race, and party dimensions, resulting in the statistics of sharing events within different states.
In each bucket of \name{}-univariate, we provide three statistics: the number of shares, the number of unique users who share the domain, and the Gini index of the sharing count across users.
And similar to \name{}-multivariate, \name{}-univariate includes four variants: monthly distribution, monthly baseline, all-time distribution, and all-time baseline.

The detailed data schema of the released versions of \name{}-multivariate and \name{}-univariate can be found in the Data Records section.

\subsection*{Derived Metrics for Domains}

Based on the sharing behavior of users in different demographic groups, we derive additional metrics that quantify different aspects of the domains.

\subsubsection*{Domain Localness Metric}

Local news is a fundamental component of the U.S. democratic process.
It is uniquely positioned to report on local affairs and elections, enabling citizens to engage in local political activities and hold their elected officials accountable~\citep{martin2019local,moskowitz2021local}.
However, the landscape of local journalism is undergoing significant changes, marked by a notable decline in local agencies, often referred to as the emergence of the ``news desert''~\citep{hayes2018decline,abernathy2018expanding}.
This trend threatens the vibrancy of local political participation and raises concerns about the overall health of democracy~\citep{rubado2020political,shaker2014dead}.

To empirically understand the dynamics of news consumption and related phenomena, it is essential to reliably categorize news outlets as either local or national.
Despite extensive research efforts, a universally accepted definition of local news organizations remains elusive~\citep{hagar2020defining}.
Many studies on local news often fail to provide a clear definition or specific criteria for classification~\citep{lowrey2008toward}, complicating efforts to expand the scope of classification and hindering the replication of analyses.

Here, we leverage the state universe data from \name{}-univariate to derive a data-driven metric that quantifies the ``localness'' of news domains.
This is achieved by calculating the deviation of the user distribution of each domain in different states from the baseline distribution, both of which are provided in \name{}-univariate.

For a formal definition, let $C_{s}$ represent the number of unique users in state $s$ across all domains, and $F_{s}$ represent the corresponding frequency, where $F_{s} = C_{s}/\sum_{s} C_{s}$.
$F_{s}$ characterizes the baseline distribution of the whole population.
For a domain $\delta$, we calculate the user frequency in state $s$, $F_{\delta, s} = C_{\delta, s}/\sum_{s} C_{\delta, s}$, where $C_{\delta, s}$ represents the number of unique users in state $s$ who share the domain $\delta$.
For domains shared by diverse users, the observed distribution $F_{\delta, s}$ should closely align with the baseline distribution $F_{s}$ across different states.
However, deviations from the baseline distribution are expected for domains with a more concentrated audience.

Following this intuition, we quantify the deviation of a domain, denoted by $\mathcal{L}_\delta$, using the Kullback–Leibler (KL) divergence between $F_{\delta, s}$ and $F_{s}$:
\begin{align}
    \mathcal{L}_\delta &= D^{(KL)} (F_{\delta, s} || F_{s}) = \sum_s F_{\delta, s} \log_2 \frac{F_{\delta, s}}{F_{s}} \text{,} \label{eq:kl}
\end{align}
where $F_{\delta, s} \log_2 (F_{\delta, s}/F_{s})$ measures the discrepancies between the observed sharing patterns and the baseline distribution of domain $\delta$ in state $s$.
$\mathcal{L}_\delta$ is a non-negative value that is minimized at zero when $F_{\delta, s}$ and $F_{s}$ are identical.
In other words, national news domains should have $\mathcal{L}_\delta$ close to zero, while local news domains should have bigger $\mathcal{L}_\delta$ values.
In the Technical Validation section, we show that $\mathcal{L}_\delta$ is a good proxy for the localness of news domains.

A limitation of $\mathcal{L}_\delta$ is that it can only indicate the deviation of a domain's sharing pattern from the baseline distribution.
To reveal which states are over-represented or under-represented, one needs to further inspect the values of $F_{\delta, s}$ and $F_{s}$.

\subsubsection*{Domain Audience Partisanship Metric}

A healthy democratic society requires the public to receive accurate and unbiased news and civic information, especially during election seasons~\citep{sunstein2006infotopia}.
However, the presence of partisan online news and phenomena such as echo chambers and filter bubbles remain concerns~\citep{cinelli2021echo,bakshy2015exposure}.
To address these issues, researchers have investigated the political biases embedded in online platforms, including search engines like Google~\citep{robertson2018auditing} and social media platforms like Facebook~\citep{bakshy2015exposure}, Twitter~\citep{chen2021neutral}, and YouTube~\citep{hosseinmardi2021examining}.
Other relevant research has focused on how users interact with different information sources and their consumption patterns~\citep{garimella2021political,robertson2023users,gonzalez2023asymmetric}.
Such analyses generally involve assessing the political leanings of numerous domains, but such datasets have been rare and often lack comprehensive coverage (see discussion in the Technical Validation section).

Here, we employ the party (and party registration) universe data from \name{}-univariate to create data-driven metrics that assess the audience partisanship of domains.
We focus on Democrats and Republicans and exclude Independent users.
The number of users from each party allows us to quantify the partisanship of the audience for each domain.
It is important to note that our audience-based metrics do not evaluate the content characteristics of these domains.
However, previous research indicates that audience characteristics are closely associated with the leanings of these domains~\citep{gentzkow2010drives,buntain2023measuring}.

Formally, the audience partisanship score $\mathcal{P}_\delta$ of a domain $\delta$ is calculated as follows:
\begin{align}
    \mathcal{P}_\delta &= \frac{
    \frac{C_{\delta, r}}{C_r} - \frac{C_{\delta, d}}{C_d}
    }{
    \frac{C_{\delta, r}}{C_r} + \frac{C_{\delta, d}}{C_d}
    }\text{,}\label{eq:bias}
\end{align}
where $C_{\delta, r}$ and $C_{\delta, d}$ (available in the distribution variant of \name{}-univariate) represent the number of unique users from the Republican and Democrat parties who share the domain $\delta$, respectively.
$C_r$ and $C_d$ (available in the baseline variant of \name{}-univariate) represent the total number of unique users in the Republican and Democrat parties who share any domain, respectively.
Since a user can share multiple domains, we have $C_{\delta, r} \le C_r \le \sum_i C_{i, r}$ and $C_{\delta, d} \le C_d \le \sum_i C_{i, d}$.
$\mathcal{P}_\delta$ is a continuous value between $-1$ and $+1$, where $-1$ means the domain $\delta$ is exclusively shared by Democratic users and $+1$ means $\delta$ is exclusively shared by Republican users.

\subsubsection*{Other Metrics}

In addition to the localness and audience partisanship metrics, we release three more audience-based metrics: age deviation, race deviation, and gender leaning, to help researchers understand the sharing patterns conditioned on these demographic variables.
The age and race deviation metrics are calculated using Eq.~\eqref{eq:kl}, where the state categories are replaced with the age or race categories.
These metrics quantify how concentrated the audience is in certain age or race groups.
The gender leaning metric is calculated using Eq.~\eqref{eq:bias}, where the party categories are replaced with the gender categories.
Similar to the audience partisanship metric, the gender leaning metric is also a continuous value between $-1$ and $+1$, where $-1$ means the domain is exclusively shared by male users and $+1$ means the domain is exclusively shared by female users.

The calculation of these metrics is very flexible.
While we primarily use the unique number of users in both Eq.~\eqref{eq:kl} and Eq.~\eqref{eq:bias}, our experiments demonstrate that using the number of shares produces highly correlated results.
The metrics can also be calculated over different time periods.
In this paper, we present results for the entire time period in the released version, case studies, and validation.
To facilitate reproducibility and customization, we provide the code for calculating these metrics, allowing readers to modify the formulas according to their specific needs.

Our formulas in Eq.~\eqref{eq:kl} and Eq.~\eqref{eq:bias} have a limitation: they rely solely on user distribution without accounting for variations in sharing patterns across demographic groups.
For example, our analysis reveals that Democratic users share more diverse domains than Republican users, averaging 74.9 unique domains compared to 54.5 across the whole period.
To enable researchers to develop more sophisticated metrics that incorporate these behavioral differences, we provide the mean number of unique domains shared by users in each demographic category and the corresponding standard deviations in the baseline variants of our datasets.

\subsubsection*{Case Studies}

\begin{figure}
\centering
\includegraphics[width=0.80\textwidth]{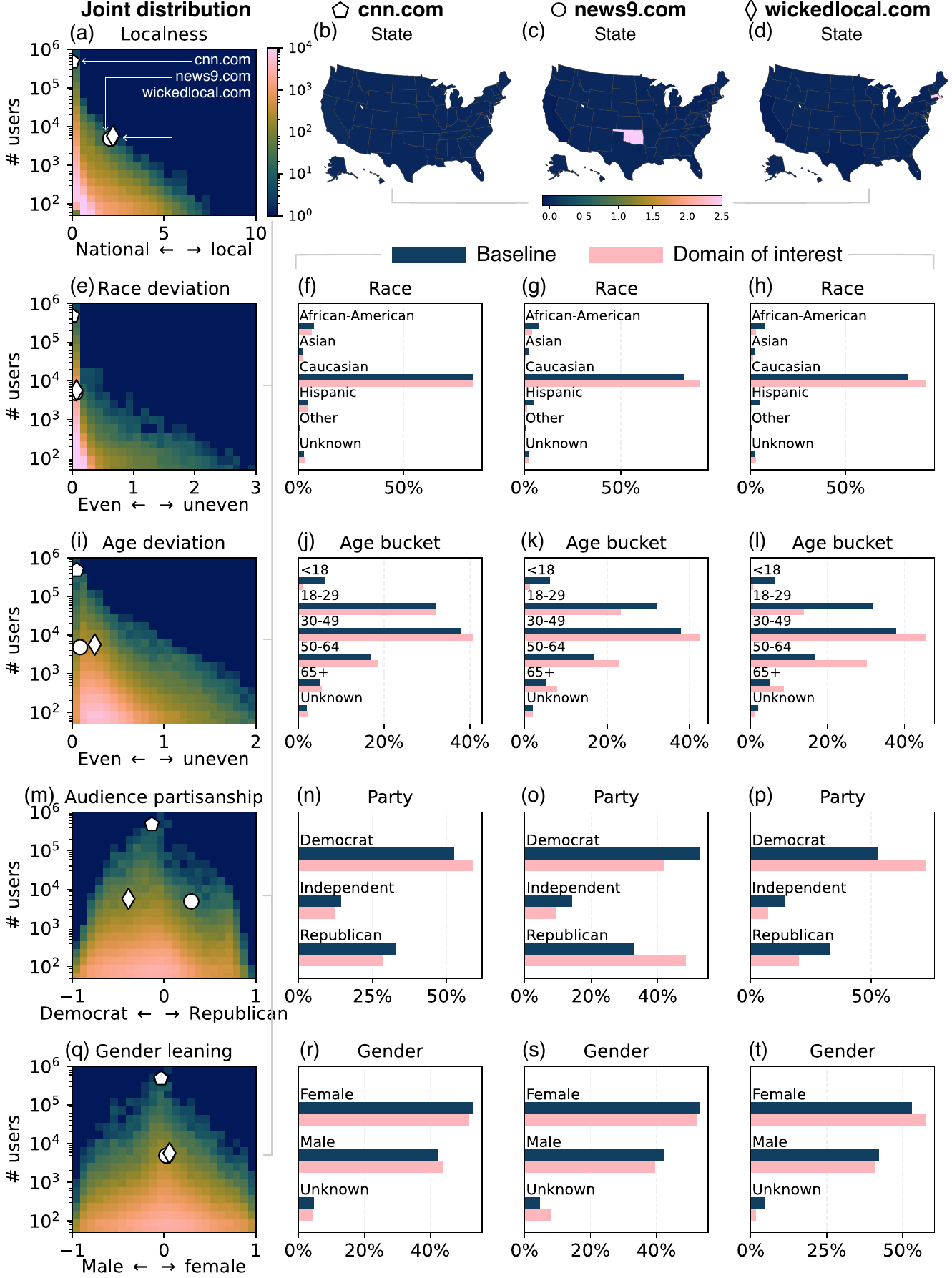}
\caption{
    Distributions of the derived metrics for all domains, along with detailed information for three domains: \domain{cnn.com}, \domain{news9.com}, and \domain{wickedlocal.com}.
    The left column presents the joint distributions of our derived metrics and the unique number of users who share each domain across the entire dataset.
    The color coding represents the number of domains within each grid cell.
    The symbols indicate the locations of the three domains.
    The three columns on the right provide detailed distributions of users in various demographic dimensions for the three domains, respectively.
    Sub-figures (b-d) highlight the discrepancies between the observed user distribution and the baseline distribution across U.S. states for the three domains.
    The color coding indicates the $F_{\delta, s} \log_2 (F_{\delta, s}/F_{s})$ value in each state.
    The bar plots display both the baseline distribution and the distribution of users in each demographic category for the domain of interest.
    The baseline distribution represents the patterns observed across all domains in the dataset.
    }
\label{fig:metric_distributions}
\end{figure}

To help the readers interpret the derived metrics, we present the distributions for all domains in the dataset and provide case studies for three example domains in Figure~\ref{fig:metric_distributions}.

Firstly, \domain{cnn.com}, a national news outlet, has a user base closely aligned with the baseline.
Consequently, its localness ($\mathcal{L}_\delta = 0.013$), race deviation ($\mathcal{L}_\delta = 0.002$), age deviation ($\mathcal{L}_\delta = 0.049$), and gender leaning ($\mathcal{P}_\delta = -0.033$) scores are near zero.
\domain{cnn.com} is shared more often by Democratic users and less often by Republican users than the baseline, resulting in an audience partisanship score of $-0.132$.

The second example, \domain{news9.com}, is a local news outlet in Oklahoma City, Oklahoma.
It is shared by fewer users than \domain{cnn.com} and has a localness score of 2.072, indicating a localized audience.
Figure~\ref{fig:metric_distributions}(c) shows that \domain{news9.com} is over-represented in Oklahoma, confirming its local nature.
Additionally, \domain{news9.com} is shared more often by Republican users and less often by Democratic users compared to the baseline, leading to an audience partisanship score of 0.297.
Its user base has race ($\mathcal{L}_\delta = 0.051$) and gender ($\mathcal{P}_\delta = 0.026$) profiles similar to the baseline but is shared more often by older users ($\mathcal{L}_\delta = 0.086$).

The third example is \domain{wickedlocal.com}, a local news source in Boston, Massachusetts.
Figure~\ref{fig:metric_distributions}(d) indicates that it is over-represented in Massachusetts, consistent with its localness score of 2.221.
Unlike \domain{news9.com}, \domain{wickedlocal.com} is shared more often by Democratic users ($\mathcal{P}_\delta = -0.387$) and even more often by older users ($\mathcal{L}_\delta = 0.246$).
Otherwise, the user base of \domain{wickedlocal.com} has a similar profile in terms of race ($\mathcal{L}_\delta = 0.071$) and gender ($\mathcal{P}_\delta = 0.060$) to that of \domain{news9.com}.

Due to space constraints, we can only provide three case studies here.
We have released an interactive app to allow readers to explore the patterns of other domains in our dataset at \url{domaindemo.info}.

\section*{Data Records}

\subsection*{Data Access}

Our dataset is available on Zenodo (\url{https://doi.org/10.5281/zenodo.15151613})~\citep{yang2025domaindemo}.
Given the sensitive nature of the information about Twitter users in our datasets, we have implemented layered access controls.
Since \name{}-multivariate and \name{}-univariate can potentially reveal the identities of Twitter users in the dataset when combined with other datasets, restrictions are imposed on access to them.
Specifically, researchers must complete an application process and sign a data use agreement that prohibits the identification of individual Twitter users and re-distribution of the data.
Those interested in accessing these datasets can follow the instructions on the Zenodo page.
The derived metrics of the domains, such as localness and audience partisanship scores, are made publicly available.

\subsection*{Data Format}

\begin{figure}
    \centering
    \includegraphics[width=\columnwidth]{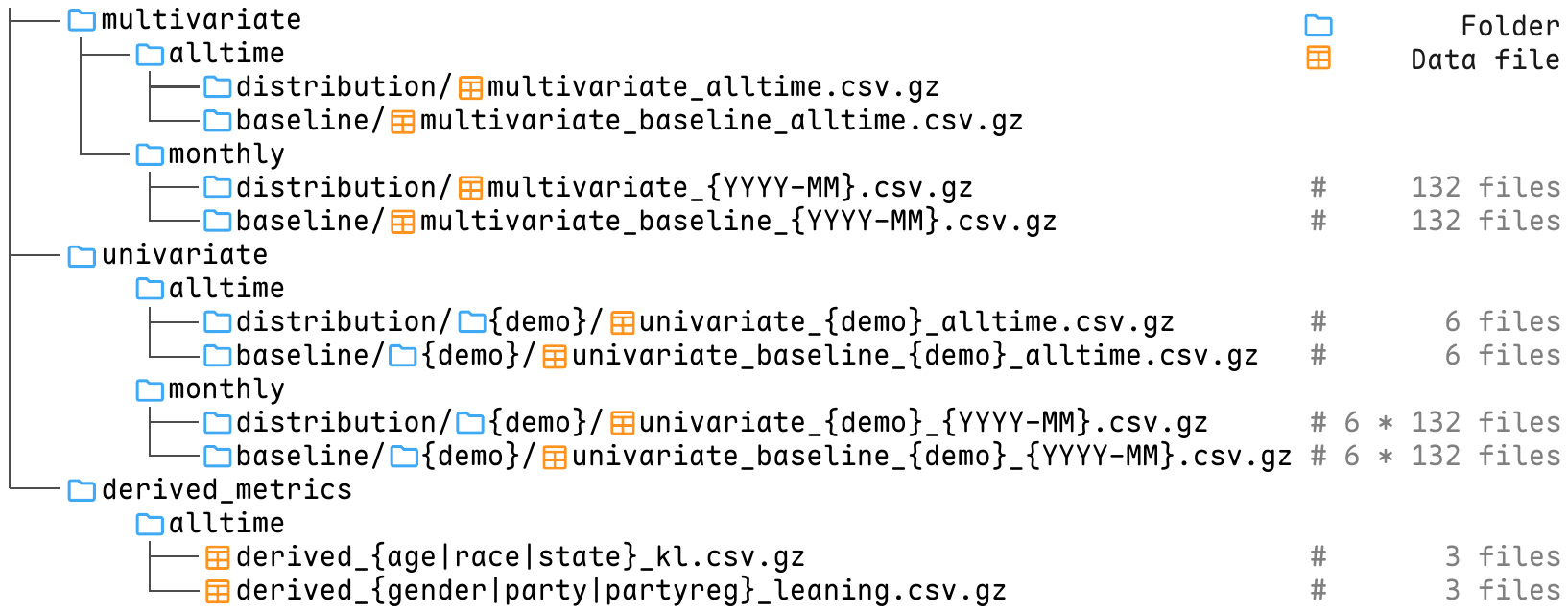}
    \caption{
        Folder structure of the DomainDemo dataset.
        We release the multivariate and univariate versions of the domain-sharing statistics.
        We provide both monthly and all-time variants of the data.
        Each variant contains distribution and baseline subdirectories.
        For clarity and due to space limitations, we use wildcards to represent patterns in file names rather than listing each file individually.
        Specifically, \{demo\} stands for different demographic factors (i.e., age, gender, race, state, party, and partyreg, six universes in total), and \{YYYY-MM\} denotes year-month combinations from May 2011 to April 2022 (132 months in total).
        We also release the derived metrics for domains based on the whole time period.
        The number of files represented by each pattern is indicated with comments on the right.
        }
    \label{fig:file_structure}
\end{figure}

Figure~\ref{fig:file_structure} illustrates the folder structure of the DomainDemo dataset.
Due to file count limitations on the data hosting platform, each root folder is distributed as a compressed archive.
After downloading and extracting these archives, users will find the subfolders and files organized according to the structure depicted in Figure~\ref{fig:file_structure}.
All data files are provided in CSV format and compressed using the Gzip algorithm for efficient storage and transmission.
Users with access can load and analyze them using preferred programming languages, such as Python and R.
In the corresponding code repository (see details in the Code Availability section), we provide example scripts to work with the data.
In the following, we provide the schema of the tables in \name{}.

\subsubsection*{Domain-sharing Statistics}

\begin{table}[t]
    \centering
    \caption{
        Schema of the \name{}-multivariate tables.
        The ``domain'' column is only included in the distribution variants, whereas the ``domains\_count\_mean'' and ``domains\_count\_std'' columns are only included in the baseline variants.
        }
    \label{tab:multivariate}
    \resizebox{\columnwidth}{!}{
    \begin{tabular}{lllp{7.5cm}}
        \hline
        Column & Type & Category & Note  \\
        \hline
        \hline
        domain & string & key & second-level domains of websites, e.g., \domain{nytimes.com}; only included in distribution variants, not in baseline variants \\
        time & string & time & year and month in the format ``YYYY-MM'' for monthly data; equal to ``alltime'' for all-time data \\
        shares & integer & statistic & sharing event count in the bucket \\
        users & integer & statistic & unique number of users in the bucket \\
        gini & float & statistic & Gini index of the sharing count across users in the bucket \\
        domains\_count\_mean & float & statistic & average number of unique domains shared by users in the bucket \\
        domains\_count\_std & float & statistic & standard deviation of the number of unique domains shared by users in the bucket \\
        state & string & demographic & two-letter abbreviations, e.g., MA; including 50 U.S. states and the District of Columbia \\
        race & string & demographic & can be one of ``African-American,'' ``Caucasian,'' ``Hispanic,'' ``Asian,''``Other,'' and ``Unknown'' \\
        gender & string & demographic & can be one of ``Male,'' ``Female,'' and ``Unknown'' \\
        age & string & demographic & age bucket; can be one of ``<18,'' ``18-29,'' ``30-49,'' ``50-64,'' ``65+,'' and ``Unknown'' \\
        party & string & demographic & can be one of ``Democrat,'' ``Independent,'' and ``Republican'' \\
        \hline
        \end{tabular}
    }
\end{table}

The table schema for \name{}-multivariate is provided in Table~\ref{tab:multivariate}.
Note that different variants of the dataset have slightly different columns, and party registration is not included.
Considering the monthly distribution variant, a row with the following values: domain=\domain{example.com}, state=CA, race=Asian, gender=Female, age=30-49, party=Democrat, year\_month=2018-12, shares=50, users=10, gini=0.1 indicates that the domain \domain{example.com} was shared 50 times in December 2018 by 10 users who live in California, are Asian females aged between 30 and 49, and identify as Democrats.
The Gini index of 0.1 indicates that the sharing count is almost evenly distributed across these users.
The baseline data, on the other hand, do not have the domain column.
A row with the following values: state=CA, race=Asian, gender=Female, age=30-49, party=Democrat, year\_month=2018-12, shares=350, users=120, gini=0.5, domains\_count\_mean=2.2, domains\_count\_std=3.5 indicates that there were 350 sharing events in December 2018 by 120 users who live in California, are Asian females aged between 30 and 49, and identify as Democrats.
These users shared an average of 2.2 unique domains with a standard deviation of 3.5.

\begin{table}
    \centering
    \caption{
        Schema of the \name{}-univariate tables. There are six universes: state, race, gender, age, party, and party registration.
        Each universe has a unique column for the corresponding demographic variable but shares other common columns.
        The ``domain'' column is only included in the distribution variants, whereas the ``domains\_count\_mean'' and ``domains\_count\_std'' columns are only included in the baseline variants.
    }
    \label{tab:univariate}
    \resizebox{\columnwidth}{!}{
    \begin{tabular}{lllp{7.5cm}}
        \hline
        Column & Type & Category & Note \\
        \hline
        \hline
        domain & string & key & second-level domains of the websites, e.g., \domain{nytimes.com}; only included in distribution variants, not in baseline variants \\
        time & string & time & year and month in the format ``YYYY-MM'' for monthly data; equal to ``alltime'' for all-time data \\
        shares & integer & statistic & sharing event count in the bucket \\
        users & integer & statistic & unique number of users in the bucket \\
        gini & float & statistic & Gini index of the sharing count across users in the bucket\\
        domains\_count\_mean & float & statistic & average number of unique domains shared by users in the bucket \\
        domains\_count\_std & float & statistic & standard deviation of the number of unique domains shared by users in the bucket \\
        state & string & demographic & two-letter abbreviations, e.g., MA; including 50 U.S. states and the District of Columbia; only included in the ``state'' universe \\
        race & string & demographic & can be one of ``African-American,'' ``Caucasian,'' ``Hispanic,'' ``Asian,''``Other,'' and ``Unknown''; only included in the ``race'' universe \\
        gender & string & demographic & can be one of ``Male,'' ``Female,'' and ``Unknown''; only included in the ``gender'' universe \\
        age & string & demographic & age bucket; can be one of ``<18,'' ``18-29,'' ``30-49,'' ``50-64,'' ``65+,'' and ``Unknown''; only included in the ``age'' universe \\
        party & string & demographic & inferred party; can be one of ``Democrat,'' ``Independent,'' and ``Republican''; only included in the ``party'' universe \\
        partyreg & string & demographic & party registration; can be one of ``Democrat,'' ``Independent,'' ``Republican,'' ``Other,'' and ``Unknown''; only included in the ``party registration'' universe \\
        \hline
        \end{tabular}
    }
\end{table}

\name{}-univariate details the statistics of sharing events across various categories within individual demographic variables.
It includes separate sets of statistics (universes) for state, race, gender, age, party, and party registration.
The table schema for these statistics is provided in Table~\ref{tab:univariate}.
A row in the state universe monthly distribution variant with the values: domain=\domain{example.com}, year\_month=2018-04, shares=3,000, users=250, gini=0.8, and state=NY indicates that the domain \domain{example.com} was shared 3,000 times in April 2018 by 250 users in New York.
The Gini index of 0.8 suggests that the sharing count is highly concentrated among a few users.
A row in the baseline variant with the values: state=NY, year\_month=2018-04, shares=451,000, users=18,250, gini=0.5, domains\_count\_mean=10.9, domains\_count\_std=10.2 indicates that there were 451,000 sharing events of any domains in April 2018 by 18,250 users in New York.
These users shared an average of 10.9 unique domains with a standard deviation of 10.2.

\subsubsection*{Derived Metrics for Domains}

\begin{table}
    \centering
    \caption{Derived metrics for domains.
    The table presents six metrics derived from six different data universes.
    From left to right, we report the metric name, size, range, data universe, calculation method, and a brief description.
    }
    \label{tab:derived}
    \resizebox{\columnwidth}{!}{
    \begin{tabular}{p{2.3cm}llllp{6cm}}
        \hline
        Metric name & Size & Range & Data universe & Calculation & Note  \\
        \hline
        \hline
        Localness & 129,127 & $[0, +\infty)$ & state & Eq.~\eqref{eq:kl} & Larger values indicate more local domains. \\
        Race deviation & 129,127 & $[0, +\infty)$ & race & Eq.~\eqref{eq:kl} & Larger values indicate that the domain shares are more concentrated on some race categories. \\
        Age deviation & 129,127 & $[0, +\infty)$ & age & Eq.~\eqref{eq:kl} & Larger values indicate that the domain shares are more concentrated on some age categories. \\
        Audience partisanship & 129,127 & $[-1, +1]$ & party & Eq.~\eqref{eq:bias} & Only including Democrat and Republican categories. Negative values indicate more shares from Democratic users and vice versa. \\
        Audience partisanship (registration) & 129,041 & $[-1, +1]$ & party registration & Eq.~\eqref{eq:bias} & Only including Democrat and Republican categories. Negative values indicate more shares from Democratic users and vice versa. \\
        Gender leaning & 129,127 & $[-1, +1]$ & gender & Eq.~\eqref{eq:bias} & Only including Male and Female categories. Negative values indicate more shares from Male users and vice versa. \\
        \hline
        \end{tabular}
    }
\end{table}

In addition to the domain-sharing statistics, we also release the derived metrics for domains based on the data from the whole time period.
Each of our released files contains two columns: domain and the corresponding metric value.
Details of these metrics are provided in Table~\ref{tab:derived}.
Note that we offer two versions of the audience partisanship scores: one based on inferred user partisanship and one based on party registration.
The audience partisanship scores based on party registration cover fewer domains than other metrics due to the missing values of the party registration information.

\section*{Technical Validation}

In this section, we discuss the robustness of the demographic variables in our dataset.
We then compare the localness and audience partisanship scores of news domains against existing classifications.

\subsection*{Demographic Variables}

The demographic variables of Twitter panel users are the foundation of DomainDemo.
While other information is self-reported, the partisanship score and race are inferred by TargetSmart.
Although the inference algorithms remain proprietary, multiple lines of evidence support their reliability.
For the partisanship score, we find that it highly correlates with the party registration information for individuals registered as Democrats or Republicans in the 30 states plus the District of Columbia where party registration information is available, with a 94\% agreement rate.
An independent evaluation from the Pew Research Center also suggests this inferred partisanship is reasonably accurate~\citep{pew2018commercial}.
Moreover, our previous research validates these scores through their strong alignment with county-level election results~\citep{shugars2021pandemics}.
Similarly, TargetSmart's race estimates show consistency with different reference points, including self-reported race data from a Pew Research Center survey and results from a statistical inference method~\citep{shugars2021pandemics}.

The representativeness of the Twitter panel is another important aspect of our dataset.
Our previous research has compared the panel with a representative sample of registered voters on Twitter created by the Pew Research Center~\citep{hughes2021using}.
The study shows substantial agreement between the two samples in general, despite some noteworthy differences.
In particular, the panel exhibits an overrepresentation of Caucasian users while underrepresenting other racial groups, particularly Hispanic and Asian populations.
Additionally, the panel contains a slightly higher share of female users and younger individuals compared to the survey samples.

\begin{figure}
    \centering
    \includegraphics[width=0.6\columnwidth]{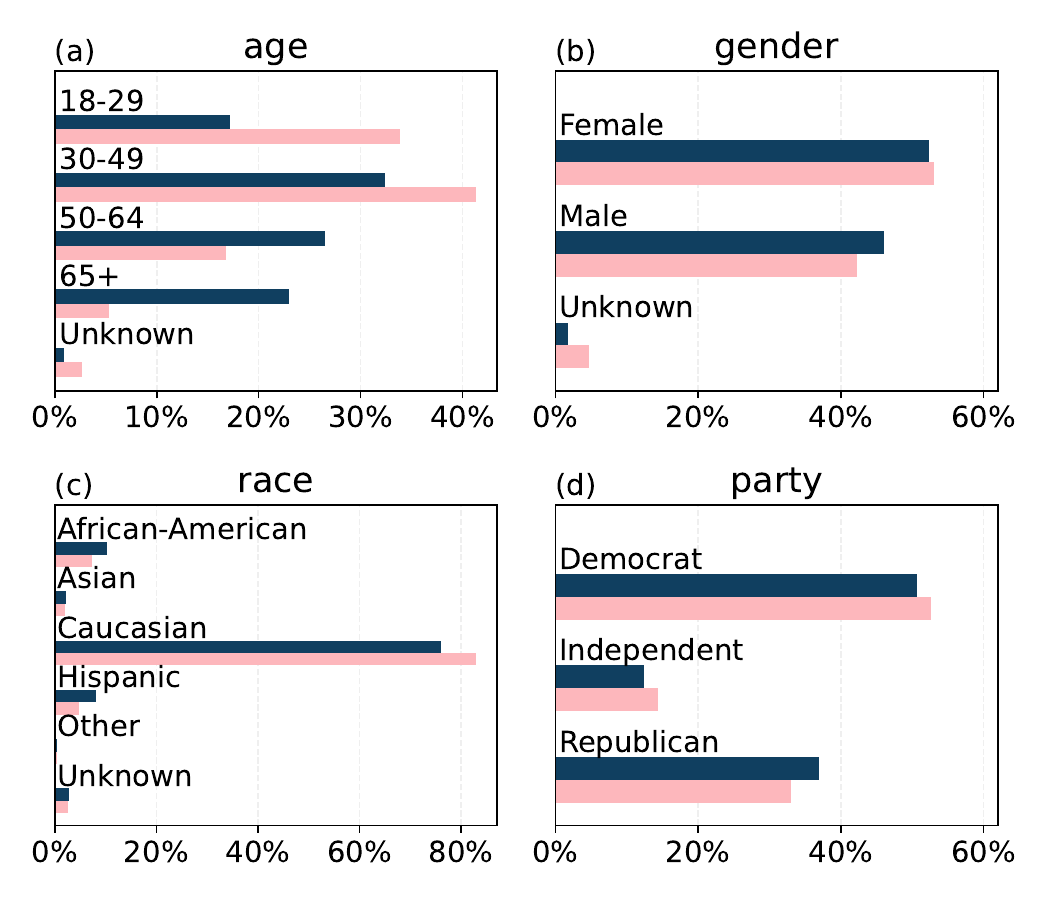}
    \caption{
        Comparison of the demographic composition of the Twitter panel with that of all registered voters.
        The age is calculated using 2017 as the reference year.
    }
    \label{fig:panel_vs_all_voters}
\end{figure}

Here, we further compare the demographic composition of the Twitter panel with all registered voters in the TargetSmart voter file, as illustrated in Figure~\ref{fig:panel_vs_all_voters}.
Notably, Twitter panel users tend to be younger than registered voters.
For other aspects, the panel generally reflects the composition of registered voters with some minor differences.
In particular, the panel contains a higher proportion of Caucasian users while underrepresenting other racial groups.
Additionally, male users and Republican users are slightly underrepresented in the Twitter panel.

These comparative analyses offer valuable insights into the representativeness of our Twitter panel.
The panel demonstrates reasonable alignment with the broader population of registered voters and voters on Twitter.
However, researchers should exercise caution when interpreting results, particularly regarding potential biases in age distribution, gender representation, and especially racial composition.

\subsection*{Domain Localness Metric}

\begin{table}
    \centering
    \caption{Summary of existing classifications of local and national news outlets. From left to right, we report the dataset, the number of local domains, the number of national domains, the total number of domains, and the description of the dataset.}
    \label{tab:existing_classification_cross_tab}
    \resizebox{\columnwidth}{!}{
    \begin{tabular}{lrrrp{7.5cm}}
        \hline
        Dataset & Local & National & All & Description \\
        \hline
        \hline
        \citet{cronin2023null} & 1,390  & 276 & 1,666 & Cronin et al. present a manually coded list of news outlets categorized as local, national, or international. We exclude international news sources from our study. \\
        \citet{fischer2020auditing} & 6,455 & 511 & 6,966 & Fischer et al. offer a classification of news domains into categories such as local, regional, national, international, and technical. Their labels are derived from multiple sources, including Yin et al. and manual annotation. For our analysis, we retain only the local and national domains. The authors do not provide specific criteria to distinguish regional from local categories and often combine them in their analysis. Therefore, we treat all regional domains as local in our study. \\
        \citet{horne2022nela} & 284 &  0 & 284 & Horne et al. compile a list of local news domains sourced from \domain{50states.com/news}. \\
        \citet{yin2018local} & 6,493 & 0 & 6,493 & Yin et al. compile a list of local news outlets. The list is assembled from multiple sources, such as \domain{usnpl.com} and \domain{stationindex.com}. \\
        abyz & 10,124 & 290 & 10,414 & \domain{abyznewslinks.com} provides a list of newspapers and news media in the U.S., classifying them as either local or national. We obtained the data from its website. \\
        \hline
        \end{tabular}
    }
\end{table}

In Figure~\ref{fig:metric_distributions}, we present the discrepancies between the observed user distribution and the baseline distribution, i.e., $F_{\delta, s} \log_2 (F_{\delta, s}/F_{s})$, across different states for three domains.
The results indicate that our localness metric can effectively capture and quantify the audience patterns of these domains.
To systematically validate our localness score, we focus on news media and compile five existing classifications of local and national news outlets.
Table~\ref{tab:existing_classification_cross_tab} provides a summary of the statistics and information for these datasets.
While these lists primarily classify news outlets based on coverage and production perspectives, our approach emphasizes the audience perspective.

We merge all these datasets into a single dataset called \dataset{meta-ln}, which contains 12,905 unique domains.
Domains are labeled as local or national when there is a consensus among the original sources.
Only 40 domains (0.31\%) have inconsistent classifications across different datasets.
These inconsistencies mainly arise from the varying definitions adopted by different authors for some borderline outlets.
For instance, \domain{abc7.com} is labeled as national by \cite{cronin2023null} but as local by other datasets.
We exclude these domains from our analysis and only keep the 4,853 news domains that are present in our dataset for further comparison.

We utilize the Area Under the Receiver Operating Characteristic Curve (AUC) score to assess the alignment between $\mathcal{L}_\delta$ and the existing labels.
The AUC score essentially measures the probability that our metric assigns higher $\mathcal{L}_\delta$ values to local domains compared to national domains (as identified by \dataset{meta-ln}).
An AUC score of 0.5 indicates random classification, while a score of 1.0 signifies perfect ranking by $\mathcal{L}_\delta$.
In our analysis, $\mathcal{L}_\delta$ achieves an AUC score of 0.983, indicating minimal discrepancies between \dataset{meta-ln} and $\mathcal{L}_\delta$.
Although our metric captures different signals than \dataset{meta-ln}, the high agreement level validates the accuracy of our localness metric and the robustness of \name{}.

In addition to \dataset{meta-ln}, we compare our localness metric with that of \citet{le2022understanding}.
Le Qu\'er\'e et al. also adopt a data-driven approach to quantify the localness of news domains from the audience perspective.
Specifically, they quantify the ``population reach'' of news domains by measuring the distance between the locations of the outlets and the users following the outlets on Twitter while accounting for the population density.
Since the population reach metric is continuous, we directly calculate its correlation with $\mathcal{L}_\delta$.
The intersection between the list shared by Le Qu\'er\'e et al. and ours has 1,342 domains and yields a Spearman correlation coefficient of 0.441 ($p < 0.001$), suggesting a moderate agreement between the two metrics.

Although $\mathcal{L}_\delta$ as a continuous value can capture nuanced differences between domains, dichotomizing the value can be beneficial in certain contexts.
For news domains, we can use \dataset{meta-ln} to establish a reasonable threshold.
By adjusting the threshold value for $\mathcal{L}_\delta$, we can compute the corresponding F1 score, which quantifies the agreement between $\mathcal{L}_\delta$ and the labels in \dataset{meta-ln}, and identify the optimal threshold that minimizes false positives and false negatives.
Our calculations indicate that a threshold of 0.243 yields the highest F1 score of 0.978.
When dealing with domains outside of \dataset{meta-ln}, researchers can first annotate a set of domains as local or national using \dataset{meta-ln}, and then use these labels to determine the optimal threshold of $\mathcal{L}_\delta$ for their specific study.

\subsection*{Domain Audience Partisanship Metric}

\begin{table}
    \centering
    \caption{
    Summary of the existing domain political leaning scores and their correlations with our audience partisanship scores.
    From left to right, we report the reference dataset, description of the dataset, number of overlapping domains, and Spearman correlation coefficients with our audience partisanship scores based on inferred partisanship (party) and party registration (party reg) information.
    All the correlation coefficients are statistically significant at the 0.001 level.
    }
    \label{tab:existing_bias_scores}
    \resizebox{\columnwidth}{!}{
    \begin{tabular}{p{4cm}p{6cm}rrr}
        \hline
        Dataset & Description & N & Party & Party reg\\
        \hline
        \hline
        \citet{bakshy2015exposure} & Audience-based scores crafted from Facebook data. & 398 & 0.940 & 0.929 \\
        \citet{eady2020news} & Media ideology scores based on Twitter data. The authors jointly estimate the ideology of politicians, users, and news sources through the news sharing behaviors on Twitter. & 179 & 0.929 & 0.929 \\
        \citet{buntain2023measuring} & Audience-based scores derived from the Facebook URL dataset. & 2,480 & 0.916 & 0.898 \\
        MBFC (mediabiasfactcheck.com) & MBFC (Media Bias Fact Check) provide rater-based political leaning categories for various news domains. We map their labels ``far-left,'' ``left,'' ``center-left,'' ``center,'' ``center-right,'' ``right,'' and ``far-right'' to numerical values $-1$, $-0.66$, $-0.33$, $0$, $0.33$, $0.66$, $1$ for analysis. &  2,986 & 0.765 & 0.774 \\
        Allsides (allsides.com/media-bias/media-bias-ratings) & Allsides produces domain bias scores based on its own algorithm. We map their labels ``left,'' ``left-lean,'' ``center,'' ``right-lean,'' and ``right'' to numerical values $-1$, $-0.5$, $0$, $0.5$, $1$ for analysis. & 189 & 0.736 & 0.743 \\
        Allsides community & Similar to the Allsides algorithmic scores, but based on crowdsourced ratings. & 189 & 0.613 & 0.611 \\
        Mturk~\citep{robertson2018auditing} & Crowdsourced ratings from Mturk. & 358 & 0.486 & 0.491 \\
        \hline
    \end{tabular}
    }
\end{table}

In a manner similar to the localness metric, we validate the audience partisanship metric by comparing it with the existing classification of domain political leaning, as detailed in Table~\ref{tab:existing_bias_scores}.
The table presents the number of domains common to both the reference dataset and our dataset, along with the Spearman correlation coefficients, all of which are positive and statistically significant at the 0.001 level.
Our ratings show a high correlation with those of \citet{bakshy2015exposure}, \citet{eady2020news}, and \citet{buntain2023measuring}, which are all audience-based scores derived from social media data.
The correlation between our metric and other existing political leaning scores that focus on the sources themselves, such as Allsides and MBFC scores, is lower, suggesting that our metric captures different signals.
Nonetheless, these findings demonstrate that our metric effectively captures the audience partisanship of various domains.

As detailed in Table~\ref{tab:derived}, we also offer a version of the audience partisanship metric derived from party registration information.
This metric shows a strong correlation with the audience partisanship scores based on inferred partisanship, exhibiting a Spearman correlation coefficient of 0.917 ($p < 0.001$) across 129,041 overlapping domains.
In Table~\ref{tab:existing_bias_scores}, we further compare this metric with existing political leaning scores.
As expected, it produces similar results to the audience partisanship scores based on inferred partisanship.

It is important to recognize that obtaining political leaning ratings for extensive sets of domains is challenging.
Most existing datasets listed in Table~\ref{tab:existing_bias_scores} cover only a few hundred domains, with a couple of them covering over 2,000 domains, primarily focusing on news outlets.
In contrast, our dataset includes scores for over 129,000 domains, encompassing a diverse array of websites beyond news sources.

By definition, our audience partisanship metric only encodes relative differences, meaning that the zero point does not necessarily indicate politically neutral.
Users seeking a binary classification could consider identifying the least biased domains in their contexts and use them to calibrate our metric~\citep{chen2021neutral}.

\section*{Usage Notes}

Our dataset offers a comprehensive view of domain sharing patterns on Twitter, capturing variations across demographic groups throughout an extended period.
The demographic characteristics of the audiences also reveal distinctive patterns that illuminate the nature of the shared domains.

A key application of our dataset lies in examining the U.S. news media landscape.
For researchers interested in this area, we refer them to a curated list of news domains~\citep{yang2025newsdomains}.
Researchers can integrate this list with our dataset to identify and analyze news domains within our collection.

Beyond news media, our dataset encompasses a diverse range of web domains.
This includes news-like websites without established editorial standards, such as misinformation sites and ``pink slime'' websites~\citep{moore2023consumption}.
The dataset also extends to various non-news domains, spanning government websites, organizational platforms, entertainment sites, and e-commerce portals.

\section*{Code Availability}

The code associated with this dataset is available at \url{https://github.com/LazerLab/DomainDemo}.
We share example scripts to load and analyze the data in \name{}.
We also provide code to reproduce the derived metrics for domains and the validation results.

\bibliography{ref}

\end{document}